\documentclass[aps,prb,reprint,superscriptaddress, floatfix]{revtex4-1} 
\usepackage{comment}
\usepackage[]{graphicx}
\usepackage{dcolumn}%
\usepackage{bm}%
\usepackage{float}%
\usepackage[utf8]{inputenc}
\usepackage{amsmath}  

\let\baraccent=\= 
\renewcommand{\=}[1]{\stackrel{#1}{=}} 

\begin{document}

\noindent \title{Shallow Trap States Control Electrical Performance of Amorphous Oxide Semiconductor Thin-Film Transistors }
\setlength\parindent{24pt}
\\
\author{Måns J. Mattsson}
\email[Corresponding author: ]{mons.mattsson@gmail.com}
\affiliation{Department of Physics, Oregon State University, Corvallis, OR, 97331-6507, USA }

\author{Jinhan Lee}
\affiliation{School of Electrical and Electronic Engineering, Yonsei University, Seoul 03722, Korea}

\author{Christopher E. Malmberg}
\affiliation{Department of Chemistry, Oregon State University, Corvallis, Oregon 97331-4003, USA}

\author{Jared Parker}
\affiliation{Department of Physics, Oregon State University, Corvallis, OR, 97331-6507, USA }

\author{Kyle T. Vogt}
\affiliation{Department of Physics, Oregon State University, Corvallis, OR, 97331-6507, USA }

\author{Hyemi Kim}
\affiliation {Advanced Device Research Lab, Samsung Electronics, Hwaseong-si, Gyeonggi-do, 18448, Korea}

\author{Minji Hong}
\affiliation {Advanced Device Research Lab, Samsung Electronics, Hwaseong-si, Gyeonggi-do, 18448, Korea}

\author{Pilsang Yun}
\affiliation {Advanced Device Research Lab, Samsung Electronics, Hwaseong-si, Gyeonggi-do, 18448, Korea}

\author{Daewon Ha}
\affiliation {Advanced Device Research Lab, Samsung Electronics, Hwaseong-si, Gyeonggi-do, 18448, Korea}

\author{Taeyoon Lee}
\affiliation{School of Electrical and Electronic Engineering, Yonsei University, Seoul 03722, Korea}

\author{Paul H.-Y. Cheong}
\affiliation{Department of Chemistry, Oregon State University, Corvallis, Oregon 97331-4003, USA}

\author{John F. Wager}
\affiliation {School of Electrical Engineering and Computer Science, Oregon State University, Corvallis, OR, 97331-5501, USA}

\author{Matt W. Graham}
\email[Corresponding author: ]{graham@physics.oregonstate.edu}
\affiliation{Department of Physics, Oregon State University, Corvallis, OR, 97331-6507, USA }




\begin{abstract}
\indent	
\\
The performance of n-type amorphous oxide semiconductor thin-film transistors (TFTs) is largely controlled by the density of states (DoS) near the conduction band mobility edge. Here, the full subgap DoS of amorphous InGaZnO (a-IGZO) TFTs, used in display panels and dynamic random-access memory (DRAM) development, is measured by ultrabroadband photoconduction (UP-DoS) microscopy to within 0.1 eV of the mobility edge. The measured subgap DoS for 25 TFT processing conditions accurately predicts each transfer curve, showing how shallow defect states are electron traps that rigidly tune subthreshold swing, threshold voltage and drift mobility. For a set of TFTs, the subthreshold transfer characteristics can be independently simulated from the experimental shallow defect DoS, with no adjustable parameters. The full transfer curve is simulated by introducing a single parameter: the conduction band tail energy. Additionally, the simulation reveals that the shallow trap density controlling subthreshold behavior can be directly extracted from transfer curves. Finally, a systematic In-enrichment study, combined with DFT+U DoS simulations, enables identification of vacancy cation coordination environments for all experimentally observed subgap peaks. The dominant trap controlling conventional a-IGZO TFT performance is centered at $\sim$0.32 eV below the conduction band mobility edge and is assigned to a Ga-Ga-In oxygen vacancy defect.
\end{abstract}
\maketitle 

\section{Introduction}
Amorphous oxide semiconductors (AOS) are an established active channel material for flat-panel display thin-film transistors (TFTs),\cite{nomura2004room,geng2023thin, kamiya2010present, bao2025amorphous} that are now being developed for next-generation dynamic random access memory (DRAM) \cite{liao2025high,ryu2024capacitorless, ha2024exploring} and neuromorphic computing architectures.\cite{jang2022amorphous, yun2025optoelectronic, chung2023visible} The amorphous and oxygen-deficient nature of AOS results in a large concentration of subgap defect states. Since the defect landscape of such TFTs often vary unpredictably with the specific processing and growth conditions, precise defect control remains an unsolved problem for AOS. \cite{mattson2023illuminating, hong2023quantitative, ide2011effects} This problem is only amplified for three-dimensional stacked DRAM architectures, where bottom transistors are known to degrade and induce new defects during the processing of subsequent layers. \cite{mao2024amorphous} While such subgap defect traps are thought to degrade charge transport properties, there is no established method that directly measures the on-chip defect density of states (DoS) to systematically predict the electrical performance of AOS TFTs.
\par  
Amorphous indium gallium zinc oxide (a-IGZO) is an n-type wide-bandgap AOS widely used in display panels and an emerging material for next-generation DRAM architectures needed for the growing data centers and AI computing demands. \cite{yan2023recent,wager2014amorphous,zhu2021indium,hu2023true} To enable innovations such as AOS DRAM, a-IGZO offers several advantages over traditional single-crystal silicon due to its low processing temperature, low leakage current, and amorphous structure, while still retaining a relatively high electron mobility. \cite{keyes2005physical,hur2024oxide,choi2024review, mao2024amorphous} 
This high electron mobility in an amorphous structure is attributed to the symmetric 5s In$^{3+}$ orbitals dominating the conduction band minimum, giving rise to a low cationic disorder. \cite{kamiya2009electronic,de2015comparison, fung2009two} The n-type doping in a-IGZO is generally attributed to a large concentration of oxygen vacancy donors. Depending on the local environment, these oxygen vacancy defect states can appear both deep and shallow within the energetic bandgap. \cite{cho2025oxygen,kumara2021local, vogt2020ultrabroadband, yoo2024direct,nakazawa2021gap} While energetically deep defect states may be related to light-induced instabilities such as negative bias illumination stressing (NBIS),\cite{zheng2025there,de2017oxygen} they are not expected to directly impact TFT transport properties. In contrast, shallow defect states, located energetically just below the conduction band mobility edge, are expected to act as electron traps that, together with conduction band tail traps, are thought to degrade TFT transport properties. \cite{hosono2022transparent,wager2022amorphous, yeon2016structural,park2022defect}
\par
To provide a first-principles approach to optimizing AOS TFTs, the first challenge is resolving the shallow defect-trap density directly on-chip.\cite{nakazawa2024reliable, vogt2020ultrabroadband}
Ultrabroadband Photoconduction Density of States microscopy (UP-DoS) is an emerging method that provides over 6 orders of magnitude sensitivity by tunable laser excitation of subgap traps directly on TFT active channel.\cite{vogt2020ultrabroadband, mattson2023illuminating, mattson2022hydrogen} Previously, UP-DoS could not optically access the shallow defects most directly impacting TFT performance. By adding a tunable difference frequency generation (DFG) laser, this work resolves this challenge by now ionizing shallow defects to within $\sim$0.1 eV of the conduction band mobility edge. 
\par
A second challenge in AOS TFT optimization lies in establishing a concrete relationship between these shallow defect states and electrical performance. Using UP-DoS to directly measure shallow defects of the active TFT channel, the experimental DoS is fed into a simulation using first-principles Fermi–Dirac statistics to model the resulting electron trapping. This new analytical approach using the TFT experimental subgap DoS may provide the missing link needed to accurately predict key TFT performance metrics directly from processing conditions. As a first step, comparison of similar TFTs processed differently demonstrates that the observed changes in shallow defects can simulate the subthreshold transfer characteristics without adjustable parameters. Then, by introducing a single adjustable parameter, the conduction band Urbach energy, the full TFT transfer curve is simulated directly from the experimental DoS of each TFT. Moreover, we show how these simulations can ultimately be inverted to instead retrieve the total shallow trap density directly from the TFT transfer curve. Lastly, by varying the indium concentration in experimental TFTs, UP-DoS suggests reasonable identities for all a-IGZO subgap oxygen vacancy peaks, with each peak identity corroborated by DFT+U defect simulations. 
\par 
  \begin{figure}
\begin{center}
   \begin{tabular}{c}
   \includegraphics[width=3.2 in.]{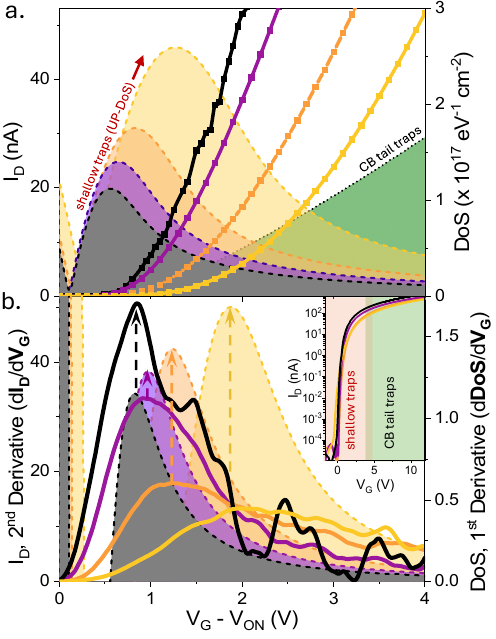}
   \end{tabular}
   \end{center}
     \caption{\textbf{ (a) } Transfer curves (solid lines) for four a-IGZO TFTs with different processing conditions. Overlaid is the independently measured shallow trap density of states (filled dashed) for each TFT. \textbf{(b) } Corresponding second derivative of the transfer curve (solid lines) overlaid with the first derivative of the experimental density of states (filled dashed) showing vertical peak alignment. In each TFT, the point of largest curvature in the transfer curve corresponds to when the subgap DoS is changing most rapidly. (\textit{inset}) Zoom-out of transfer curves with the different operating regions shaded.
    } 
     \label{fig:1}
\end{figure}

  \begin{figure} [h]
\begin{center}
   \begin{tabular}{c}
   \includegraphics[width=3.2 in.]{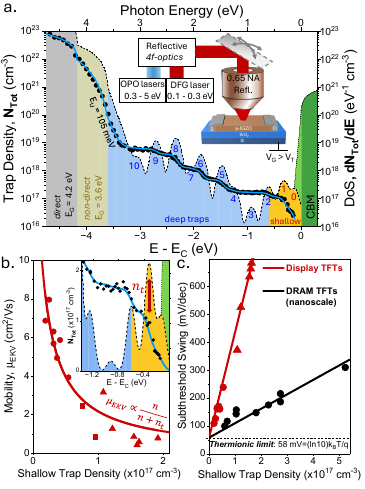}
   \end{tabular}
   \end{center}
     \caption{  \textbf{(a)} The ultrabroadband photoconduction spectrum (black dots) and the corresponding DoS (first derivative, dashed line) for a representative a-IGZO TFT is characterized by three distinct regions:  i) direct bandgap (gray), (ii) non-direct bandgap (gold), (iii) subgap defects (blue and yellow) composed of an Urbach tail and Gaussian peaks numbered 0-10. \textit{(inset)} Diagram of the UP-DoS experimental method.  \textbf{(b)} Mobility, $\mu_{EKV}$, as a function of experimentally measured shallow trap density (see inset) for 15 different a-IGZO TFTs. \textbf{(c)} Subthreshold Swing plotted against the experimentally measured shallow trap density for TFTs designed for both display (red) and DRAM (black) applications.}  
     \label{fig:2}
\end{figure}
\section{Results and Discussion}
Figure \ref{fig:1}a shows a set of four transfer curves (solid lines) plotted on a linear scale as a function of gate overvoltage (I$_{D}$-(V$_{G}$ - V$_{ON}$)). These four curves correspond to four a-IGZO TFTs fabricated under different growth and processing conditions. Overlaid is the shallow trap density of states (filled dashed) for each TFT that is independently obtained using ultrabroadband photoconduction density of states (UP-DoS) microscopy that uses tunable mid-IR lasers to directly observe the filled shallow defect states. Figure \ref{fig:1}b is a corresponding plot of the second derivative of the transfer curves ($\frac{d^2I}{d^2V_G}$) and the first derivative of the density of states (dDoS/d$V_G$). The inset shows the four transfer curves on logarithmic (I$_{D}$ - V$_G$) scales.
\par
Figure \ref{fig:1}b shows a striking correlation between each respective second derivative curve and the shallow trap density of states derivative; namely, both curves peak at the same overvoltage. Specifically, this shows the curvature of the transfer curve is maximal when the shallow trap density of states changes most rapidly. The best performing TFT (black) peaks at an overvoltage of $\sim$1 V while the worst performing TFT peaks at $\sim$ 2 V. This peak shift indicates that the poorer TFT performance is a result of more electron trapping, which persists over a larger overvoltage. Additionally, TFT performance correlates with shallow trap density as witnessed by the much larger magnitude of the yellow curve compared to the black curve of Figure \ref{fig:1}b. The transition between subthreshold and linear behavior occurs approximately when the conduction band tail state density exceeds the shallow trap density, which occurs at an overvoltage of $\sim$1.5 V ($\sim2.5$ V) for the black (yellow) curve.
\par
Although a-IGZO TFT performance depends primarily on the most shallow traps and conduction band tail state densities, a wealth of information on the subgap trap density is available from UP-DoS analysis, as indicated in Figure \ref{fig:2}a. In addition to the shallow traps shown in yellow and labeled as number Peak 0 and 1, nine other Gaussian-shaped deep traps (blue) are typically observed in a-IGZO. Furthermore, for the representative a-IGZO TFT shown in \ref{fig:2}a, a direct bandgap of 4.2 eV, an optical (non-direct) bandgap of 3.6 eV, and a valence band Urbach energy of 105 meV are extracted and labeled. Depending on processing conditions, optical bandgaps of a-IGZO are typically measured by UP-DoS to be in between 3.0 and 3.6 eV (see Supplementary Information Section S2). As shaded yellow in the inset of Figure \ref{fig:2}b, the two most shallow traps and a portion of the third most shallow trap are those responsible for establishing TFT mobility and switching performance. This is shown by the data of Figure \ref{fig:2}b in which the EKV mobility is found to degrade dramatically with increasing shallow trap density in a-IGZO display TFTs. This degradation follows the inverse dependence predicted by the trap-limited mobility model (see Equation \ref{eqn:trap}), \cite{nenashev2019percolation,wager2022amorphous} with the red solid line showing the best fit to this model.
Figure \ref{fig:2}c presents the trend in subthreshold slope with increasing shallow trap densities for two types of TFTs, i.e, display TFTs and nanoscale DRAM TFTs. For both TFT device architectures, the relationship between subthreshold swing and trap density is in excellent agreement with a linear fit constrained by the thermionic limit, ln(10)k$_B$T/q. For details regarding the construction of Figures \ref{fig:2}b and c, and the mobility correlation in nanoscale DRAM TFTs, see Supplementary Information section S5.

\begin{figure}[!htb]
    \centering
    \includegraphics[width=3 in.]{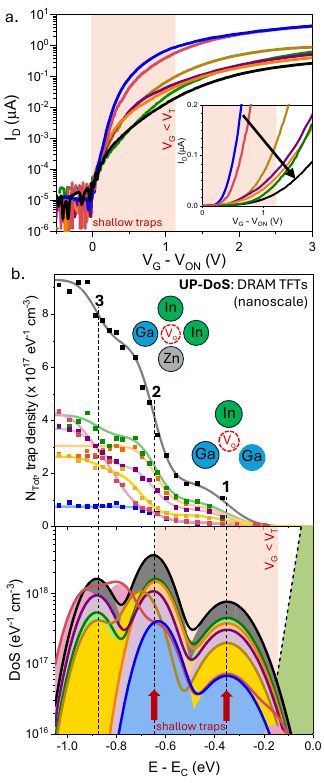}
 \caption{ \textbf{(a)} Transfer curves of 7 different nanoscale DRAM a-IGZO TFTs processed under different growth and annealing conditions.  \textbf{(b)} The ultrabroadband photoconduction DoS spectra near the conduction band mobility edge for the 7 different TFTs showing distinct steplike features. \textbf{(lower panel)} The corresponding experimentally measured subgap density of states for each TFT. The shallow trap density, highlighted by the red shaded region, is observed to be directly correlated to the TFT performance.  }
    \label{fig:Samsung}
\end{figure}
Figure \ref{fig:Samsung} presents a subset of transfer curves and UP-DoS data of the DRAM TFTs used to construct Figure \ref{fig:2}c. The I$_{D}$-(V$_{G}$ - V$_{ON}$) transfer curves included in Figure \ref{fig:Samsung}a are of dramatically different quality. Figure \ref{fig:Samsung}b demonstrate that a-IGZO TFT performance depend strongly on trap density (upper panel) and corresponding subgap DoS (lower panel). The best performing (blue) TFT transfer curve has a much smaller trap density than that of the worst (black) TFT. It turns out that a-IGZO switching and mobility performance depends exclusively on traps within $\sim$0.65 eV of the conduction band mobility edge, as evident by the comparatively well performing red curve with a large deep trap density.  As the shallow trap density increases in \ref{fig:Samsung}b, a larger gate voltage is required to fill these traps and to modulate the quasi-Fermi level out of the subthreshold into the linear regime of TFT operation. This larger shallow trap density increases the subthreshold slope and threshold voltage, and decreases the drift mobility of the TFT.
\par
As shown in the lower panel of Figure \ref{fig:Samsung}b, three main Gaussian-shaped traps are experimentally observed in this energy range. Only the two traps closest to the conduction band mobility edge (red shading) are relevant to a-IGZO TFT electrical operation. DFT+U assessment (see Section \ref{section DFT}) assigns all three peaks to oxygen vacancy donors. Moreover, peaks 1 and 2 are ascribed to a Ga-rich, 3 -atom and a In-rich, 4-atom cation coordination around a missing oxygen, respectively, as sketched in Figure \ref{fig:Samsung}b.

\subsection{Simulation of Transfer Curves from Experimental DoS}

\begin{figure*}[!htb]
    \centering
    \includegraphics[width=1.0 \linewidth]{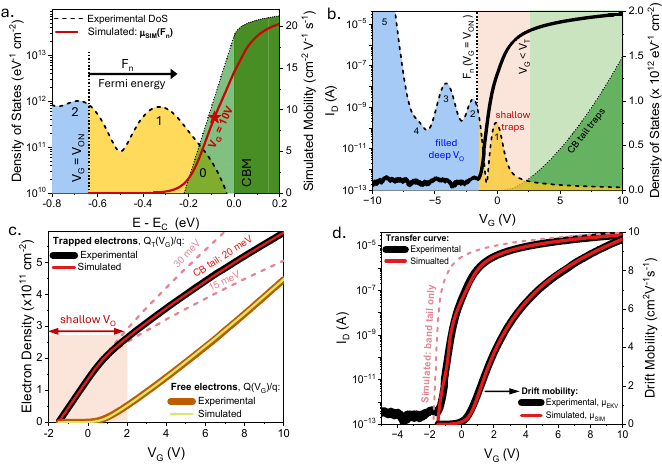}
    \caption{ \textbf{(a)} TFT simulated mobility ($\mu_{SIM}$, red line) as a function of quasi-Fermi level energy. $\mu_{SIM}$ is obtained by modeling trapping in experimentally measured Gaussian-like subgap states (filled dashed) and conduction band tail states (filled green). \textbf{(b)} a-IGZO TFT transfer curve (black, left y-axis) plotted with the experimentally measured subgap defect DoS (filled dashed, right y-axis) converted to a gate voltage axis, showing the dominant electron traps for different TFT operating regions. \textbf{(c)} Experimentally extracted (thick lines) and DoS simulated (thin lines) free and trapped electron density as a function of gate voltage. \textbf{(d)} DoS simulated transfer and mobility curves (red) plotted with experimentally measured data (black). Red dashed line is simulated transfer curve if Gaussian trap contributions are excluded. }
    \label{fig:simulation}
\end{figure*}
Figure \ref{fig:simulation}a plots the experimentally measured subgap DoS (filled dashed), and the corresponding simulated mobility when the quasi-Fermi level energy, F$_n$, is modulated from -0.64 eV to 0.2 eV (red line, right y-axis) relative to the conduction band mobility edge. Using the full DoS model shown in Figure \ref{fig:simulation}a, the trapped electron density, n$_T$(F$_n$) = Q$_T$(F$_n$)/q, and free electron density, n(F$_n$) = Q(F$_n$)/q, are calculated using Fermi-Dirac statistics (see Section \ref{sec:sim} for details). From Q$_T$(F$_n$)/q and Q(F$_n$)/q, the simulated TFT drift mobility, $\mu_{SIM}(F_n)$, is subsequently obtained from Equation \ref{eqn:trap}. When the quasi-Fermi energy is positioned 0.64 eV below the conduction band mobility edge (horizontal dotted line), essentially all electrons induced by gate overvoltage are trapped, resulting in a very small drift electron mobility. Conversely, for V$_G$ = 10 V, the quasi-Fermi level energy is simulated to be within $\sim 80$ meV from the conduction band mobility edge, yielding a corresponding drift mobility of $\sim$10 cm$^2$ $V^{-1}$ $s^{-1}$. 
\par
To better visualize electron trapping in a-IGZO TFTs,  Figure \ref{fig:simulation}b plots the experimentally measured subgap DoS (filled dashed, right y-axis) on a gate voltage axis in order to directly compare the relevant trap distribution to the experimental TFT transfer curve (solid line, left y-axis). More specifically, the dashed line represents the subgap DoS at the given quasi-Fermi level energy, which is directly mapped to a gate voltage axis using the charge sheet approximation model (see Section \ref{sec:sim}).
At the smallest gate overvoltages, Figure \ref{fig:simulation}b shows that the oxygen vacancy states, peaks 1 and 2, dominate the subgap DoS, such that most electrons induced into the channel are trapped by these states. As the gate voltage increases, the quasi-Fermi level moves towards the conduction band, filling more oxygen vacancy (V$_O$) trap states. This quasi-Fermi level modulation result in an exponentially increasing fraction of electrons populating extended states as free carriers. The larger the shallow V$_O$ trap density, the more sluggish the exponential turn on, as observed by Figure \ref{fig:2}c. When the quasi-Fermi level is modulated into the subgap DoS region that is dominated by conduction band tail states, most shallow V$_O$ trap states become filled, causing the TFT to transition out of subthreshold and enter the linear operating regime (V$_G$ $>$ V$_T$).
\par
Applying the DoS model outlined in Figure \ref{fig:simulation}a, Figure \ref{fig:simulation}c presents a simulation of the trapped electron density (Q$_T$(V$_G$)/q, red line), and the free electron density (Q(V$_G$)/q, yellow line) induced into the channel by the applied gate overvoltage. The simulated trends agree well with the experimentally extracted gate voltage-induced electron densities (black and brown lines) obtained from the TFT transfer curve. In the subhthreshold region, as a consequence of the Gaussian shallow V$_O$ trap peaks, Q$_T$(V$_G$)/q increases linearly with gate voltage and Q(V$_G$)/q increases exponentially. Above threshold (V$_G>$V$_T$), Figure \ref{fig:simulation}c reveals that both Q$_T$/q and Q/q increase linearly. The change of dominant active trap, from Gaussian-shaped V$_O$ (subthreshold) to the exponential conduction band tail (above threshold), is evident by the change of slope in Q$_T$(V$_G$)/q which occurs at $V_G \approx$ V$_T$. In the subthreshold operating regime, since virtually all states induced into the channel by applied gate voltage is trapped in Gaussian shaped V$_O$, the slope of Q$_T$(V$_G$)/q is simply C$_{ox}$/q. Above threshold, the slope of the line is C$_{ox}$/q multiplied with the fraction of carriers trapped in the exponential band tail state, which depends on the exponential slope parameter, W$_{TA}$. To fit this slope of the experimental data, a conduction band tail Urbach energy W$_{TA}$ = 20 meV is required. We consider this value to be a reliable estimate of the conduction band tail Urbach energy for this a-IGZO TFT. However, it is important to note that while shallow V$_O$ traps dominate the subthreshold region, significant above threshold non-idealities, such as series resistance or fringe current effects, \cite{kim2025mobility} may affect the W$_{TA}$ estimation by this method. Both series resistance and fringe current artifacts are minimized in this example by choosing a large-area TFT with a large W/L ratio (W = 200 $\mu$m, L = 10 $\mu$m).
\par
To complete this simulation using experimental DoS, the simulated gate voltage induced electron densities, Q$_T$(V$_G$)/q and Q(V$_G$)/q, are employed to determine the corresponding TFT drift mobility, $\mu_{\mathrm{SIM}}$(V$_G$), and drain current, $I_{D,\mathrm{SIM}}$(V$_G$). Figure \ref{fig:simulation}d shows remarkable agreement between the final simulated (red line) and the overlaid experimental transfer curve. Likewise, the simulated drift mobility curve (red line, right y-axis) show excellent agreement with the overlaid experimental mobility curve. The red-dashed line plots a simulated transfer curve if the conduction band tail is the only source of trapping, showing idealized behavior of an a-IGZO TFT without shallow V$_O$ traps. Collectively, the above method accurately simulates the transfer curve directly from experimentally measured subgap DoS, and demonstrates that both conduction band tail and shallow Gaussian trap states are required in order to simulate transfer curve behavior in an a-IGZO TFT.   
\begin{figure*}
    \centering
    \includegraphics[width=1 \linewidth]{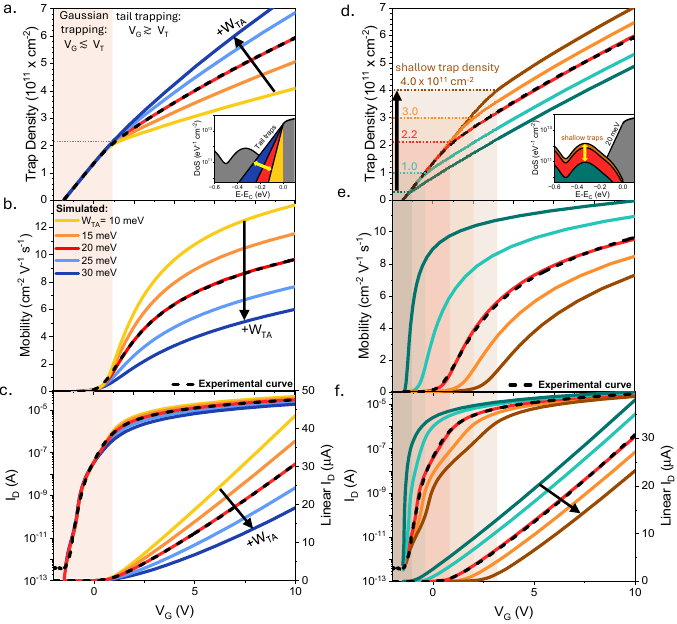}
    \caption{ DoS simulated \textbf{(a)} TFT gate voltage induced trap density, \textbf{(b)} mobility, and \textbf{(c)} transfer curve as a function of gate voltage for different conduction band tail state Urbach energies. The slope of the transfer curve in the linear regime increases with Urbach energy, while the subthreshold behavior is unaffected. DoS simulated TFT \textbf{(d)} gate voltage induced trap density, \textbf{(e)} mobility, and \textbf{(f)} transfer curve for different shallow V$_O$ trap densities. With increasing shallow V$_O$ trap density, the threshold voltage and subthreshold swing increase as more trap states need to be filled before the TFT can enter the linear regime.} 
\label{fig:simulation2}
\end{figure*}
\par
Presented in Figure \ref{fig:simulation2} are simulated gate voltage-dependent (a) trap density, (b) mobility, and (c) transfer curves for varying conduction band Urbach energies of 10 to 30 meV. In these simulations, V$_{ON}$ is set to -1.5 V in order to match the experimental TFT curve (black dashed). While subthreshold region behavior (shaded in red) largely remains constant, enhanced trapping at large gate voltages occurs with increasing Urbach energy (Figure \ref{fig:simulation2}a), resulting in a reduced mobility and a corresponding reduced current (Figure \ref{fig:simulation2}b and c). Above threshold, the simulated value of W$_{TA}$ is directly related to the slope of both Q$_T$(V$_G$)/q and the corresponding transfer curve. This implies that for an idealized TFT, the only two parameters that increase TFT  drift mobility in the linear operating regime are the conduction band tail Urbach Energy W$_{TA}$ and the trap-free mobility $\mu_o$.
\par
In Figure \ref{fig:simulation2}, the simulated (d) trap density, (e) mobility and (f) transfer curve are plotted when varying the total shallow V$_O$ trap density, from 4 to 0.25 $\times$ 10$^{11}$ cm$^{-2}$. 
The colored shaded region marks the subthreshold region for each simulated V$_O$ shallow trap density. The threshold voltage increases with increasing trap density due to more trapping prior to the TFT entering the linear regime.  In Figure \ref{fig:simulation2}d, the first linear section of the gate induced trap density (subthreshold) extends with increasing V$_O$ shallow traps, while the second linear section (above threshold) exhibits a corresponding linear offset.
Notably, the slope of both linear sections remains constant for each shallow trap density. Comparing Figures \ref{fig:simulation2}a and \ref{fig:simulation2}d, it is evident that the induced trap density as a function of gate voltage is characterized by two linear functions. The first linear function involves shallow V$_O$ trapping in the subthreshold regime, where the amount of traps controls the extension of the line along the gate voltage axis. The second line, corresponding to the above threshold regime, is due to the conduction band tail trapping, where the slope of the line is controlled by the Urbach energy and where the offset is controlled by shallow V$_O$ trap density. Consequently, as indicated by the dotted lines in Figure \ref{fig:simulation2}d, for each Q$_T$(V$_G$)/q curve, the shallow V$_O$ trap density is effectively equal to the gate voltage induced trap density at the point of slope change. Remarkably, since Q$_T$(V$_G$)/q curves can be directly obtained from experimental transfer curves (see Section \ref{sec:sim}), the total shallow V$_O$ trap density in a TFT can be simply estimated without requiring use of a defect characterization technique such as UP-DoS. The conduction band tail Urbach energy can likewise be estimated from the slope of the second linear segment of the Q$_T$(V$_G$)/q curve. However, in contrast to V$_O$ shallow donor estimation, the accuracy of Urbach Energy estimation depends on precisely knowing $\mu_0$ and on minimizing series resistance and peripheral current artifacts. 

\par
In Figure \ref{fig:simulation2}e the TFT drift mobility decreases and the threshold voltage increases with increasing shallow V$_O$ trap density as more electrons are being trapped. These mobility curves are stretched and increasingly distorted along the gate voltage axis with increasing shallow trap density. Corresponding transfer curves, as shown in \ref{fig:simulation2}f, are also stretched and distorted along the gate voltage axis. Figure \ref{fig:simulation2}f shows the evolution of a kink in the transfer curve that originates from enhanced trapping and sluggish modulation of the Quasi-Fermi level. Comparing Figure \ref{fig:simulation2}c to Figure \ref{fig:simulation2}f suggests that this distortion (or kink) observed in the simulated transfer curve is a result from a high V$_O$ shallow trap density. Finally, it is important to note that the transfer curves can only be accurately simulated when the specific peak energies and widths measured by UP-DoS are used. For examples of how the transfer curve is expected to be modulated by a different Gaussian peak energy and width, see Supplementary Information S6.
\label{Apple}
\begin{figure*}
    \includegraphics[width=7 in. ]{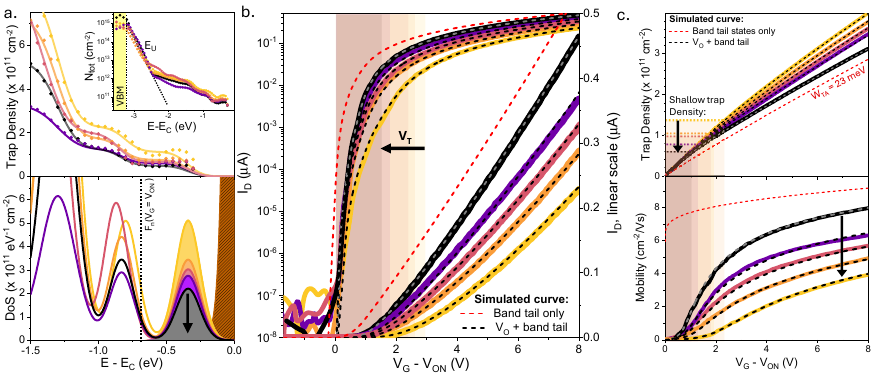}
    \caption{  \textbf{(a)} Measured shallow trap density (upper panel) for 5 a-IGZO TFTs with different processing conditions and (lower panel) corresponding subgap DoS revealing one shallow oxygen vacancy defect peak. Inset displays the full UP-DoS trap density. \textbf{(b)} Experimental TFT transfer curves (solid lines) compared to simulated transfer curves (dashed lines) generated using measured shallow subgap DoS. \textbf{(c)} Experimental (solid lines) and DoS simulated (dashed lines) gate voltage induced trap density (upper panel) and drift mobility (lower panel)
    \label{fig:Apple}
    }
\end{figure*}

\begin{center}
\begin{table*}
 \begin{tabular}{c c c c c} 
Line & V$_O$ Peak Density & S.S & Mobility, $\mu_{EKV}$& Simulated W$_{TA}$ \\
 Color & $\mathrm{\times 10^{11}}$ $(\mathrm{cm^{-2} {eV^{-1}}})$ & (mV/dec) & (cm$^2$ $V^{-1}$ $s^{-1}$) & (meV) \\

 \hline

Black & $\mathrm{2.2}$ & $\mathrm{126}$ & $\mathrm{8.0}$ & $\mathrm{23}$\\

Purple & $\mathrm{2.8}$ & $\mathrm{165}$ & $\mathrm{6.3}$ & $\mathrm{26}$\\

Red & $\mathrm{3}$ & $\mathrm{161}$ & $\mathrm{5.7}$ & $\mathrm{28}$\\

Orange & $\mathrm{3.5}$ & $\mathrm{170}$ & $\mathrm{4.9}$ & $\mathrm{30}$\\

Yellow & $\mathrm{5.1}$ & $\mathrm{238}$ & $\mathrm{3.9}$ & $\mathrm{31}$\\

 \hline
 \end{tabular}
  \caption[example]
 { \label{table:apple} Experimentally measured shallow V$_O$ peak trap density for five a-IGZO TFTs with different processing conditions compared to the corresponding electrical characteristics and simulated conduction band tail Urbach energy.}
 \end{table*}
\end{center}
\par The upper panel of Figure \ref{fig:Apple}a shows a plot of the shallow trap density and trap density across the full band gap (inset) for 5 different a-IGZO TFTs developed for display panel applications. The bottom panel of Figure \ref{fig:Apple}a plots corresponding subgap DoS, revealing one shallow oxygen vacancy defect peak.
Figure \ref{fig:Apple}b presents the comparison between experimentally measured transfer curves (solid lines) and DoS simulated transfer curves (dashed lines). The red dashed line simulates an idealized TFT without shallow oxygen vacancy traps. Black (and gray) dashed lines are simulated transfer curves by using the shallow oxygen vacancy DoS experimentally measured in each TFT and show excellent agreement with each corresponding experimental curve. Remarkably, the only adjustable simulation parameter in the Figure \ref{fig:Apple}b simulated transfer curves is the conduction band tail Urbach energy, W$_{TA}$ = 23 - 31 meV, which is still needed to modulate the slope of the linear operating regime. Figure \ref{fig:Apple}b shows robust agreement is achieved between experiment and simulation only when both shallow trap filling (subthreshold, from UP-DoS) and conduction band tail trap filling (above threshold) are included. The simulated transfer curves with no adjustable parameters (i.e., assuming fixed W$_{TA}$ for all TFT processing conditions) also agree well with experimental curves until the transition into the linear operating regime (see Supplementary Information Section S8). As such, the specific curvature of the subthreshold region, as well as characteristics such as subthreshold swing and threshold voltage (V$_T$ - V$_{ON}$), which in turn also pinpoint the onset of the linear operating regime, are well simulated from the experimentally measured subgap DoS. The simulated conduction band tail Urbach energy, along with the measured shallow trap density and electrical characteristics for each TFT, are summarized in Table \ref{table:apple}, with the specific processing conditions additionally listed in Supplementary Table S2. 
\par
Figure \ref{fig:Apple}c upper panel plots the extracted (solid lines) and DoS simulated (dashed lines) trap densities, Q$_T$(V$_G$)/q, as a function of gate voltage for the 5 different TFTs. By including the measured experimental shallow V$_O$ density in each TFT simulation, an excellent agreement between experimental and simulated results is evident. In accordance with Figure \ref{fig:simulation2}d, the shallow oxygen vacancy trap density can be extracted by the total trap density value at the point of slope change in Q$_T$(V$_G$)/q curve for each TFT. Precisely, this point of slope change occurs at the gate voltage where the 2nd derivative of the transfer curve peaks (see Figure \ref{fig:1}b). Note that, in the idealized DoS simulation (red dashed), no point of slope change occurs as the shallow trap density is zero. For large gate voltages, the extracted trap densities diverge from each other owing to an increasing W$_{TA}$, which augments the corresponding transfer curves' slope in the linear operating regime. Figure \ref{fig:Apple}c lower panel plots simulated (dashed lines) and measured mobility (solid line) curves for each TFT. In agreement with simulated trends in Figure \ref{fig:simulation2}e, the experimentally observed increase in shallow oxygen vacancy trap density results in a reduction in TFT drift mobility and an associated increase in threshold voltage, V$_{T}$ - V$_{ON}$.  \begin{figure*}
    \centering
    \includegraphics[width=7 in.]{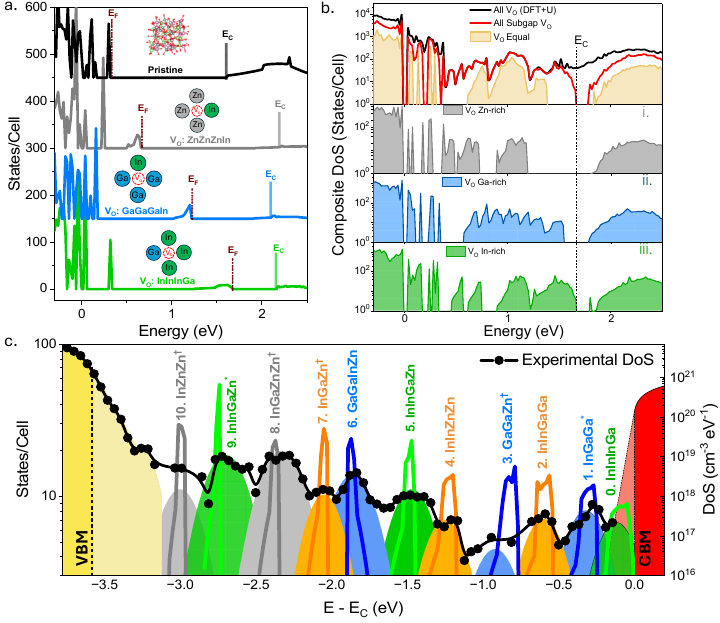}
    \caption{\textbf{(a)} Representative of DFT+U–calculated density of states (DoS) for a pristine unit cell (black line) and for cells with Zn-Zn–Zn–In (gray line), Ga–Ga–Ga–In (blue line), and In–In–In–Ga (green line) metal coordinated oxygen vacancies (V$_O$). \textbf{(b)} DFT+U calculated total summation of DoS for different oxygen vacancy configuration: all DFT+U (black), all subgap V$_O$ (red), equal V$_O$ cation coordination (beige), Zn-rich V$_O$ (gray), Ga-rich V$_O$ (blue) and In-rich V$_O$ (green). \textbf{(c)} Experimentally measured subgap DoS for a-IGZO (black dots), together with representative DFT+U-calculated oxygen vacancy defect states attributed to the experimental defect peaks (colored lines).} 
    \label{fig:DFT}
\end{figure*}
\subsection{Experimental DoS Peak Correspondence with DFT+U Assignments}\label{section DFT} 
The objective of the remaining two sections is to identify the microscopic origin of the experimentally observed subgap defect states. Figure \ref{fig:DFT}a presents the DFT+U simulated DoS for a pristine (no oxygen vacancies) a-IGZO unit cell (black) and for cells possessing oxygen vacancies with ZnZnZnIn (gray), GaGaGaIn (blue) and InInInGa (green) nearest-neighbor coordination environments. The valence band maximum defines the reference energy (VBM = 0 eV). The Fermi level energy (dotted lines) and conduction band minimum (solid lines) are marked in each simulation. The specific oxygen vacancy coordination environments shown in \ref{fig:DFT}a are selected as an example from a total of 160 simulated vacancy configurations (80 per unit-cell structure). For all simulated cells, including the pristine cell, localized states appear near the valence band maximum and are attributed to the bond length disorder on the anion (oxygen) sublattice. For the pristine case, the Fermi level energy is positioned just above these localized donor-like states. 
\par
In contrast, for all cells containing an oxygen vacancy, the Fermi level is shifted towards the conduction band and is positioned above a new defect state created within the bandgap. This new oxygen vacancy defect state appears deep for the Zn-rich oxygen vacancy cell, near the midgap for the Ga-rich oxygen vacancy cell, and shallow for the In-rich oxygen vacancy-metal coordination environment. The subgap oxygen vacancy defect state appears in all DFT+U simulations in which the neighboring cation coordination relaxes inward towards the vacancy. When neighboring coordinated cations relax outward from the vacancy, the associated defect peak appears within the conduction band, with its corresponding Fermi level positioned above the conduction band minimum (see Supplementary Information Section S3). For all simulations featuring a subgap oxygen vacancy state, the DFT+U calculated conduction band minimum is shifted to higher energies, increasing the bandgap relative to that of the pristine (oxygen defect-free) cell. This is in contrast to the case of an oxygen vacancy defect state appearing above the conduction band minimum, in which the approximate bandgap of the pristine cell is retained. 
\par
Since an oxygen atom is removed at random in all simulations, it is reasonable to sum over each calculated DoS for one cell structure to observe statistical trends. Figure \ref{fig:DFT}b (upper panel) plots the total simulated DFT+U DoS as summed over for all 80 simulations (black line) and for all subgap oxygen vacancies (red line). Additionally, all oxygen vacancies with no sole dominant metal coordination (1:1:1 triple and 2:2 quadruple vacancy-cation coordination, e.g., InGaZn or InInZnZn) are plotted in filled beige. When all simulations are summed together, the defect states positioned near the valence band, ascribed to anion sublattice bond length disorder, give rise to a pronounced smearing of the valence band. This valence band smearing is experimentally witnessed as an exponential valence band Urbach tail (W$_{TA}$ $\approx$ 100 meV) and as the emergence of an optical (non-direct) bandgap of the material (E$_g$ $\approx$ 3.0-3.6) eV.
\cite{mott2012electronic,klein2023limitations,chen2015substrate,shangguan2024review}
\par 
Figure \ref{fig:DFT}b (lower panels) plots DoS as summed over cells with oxygen vacancies in Zn- (gray), Ga- (blue), or In- (green) dominated coordination environments. Oxygen vacancy sites with Zn-rich metal coordination environments exclusively appear in the lower portion of the bandgap. Consequently, Zn-rich oxygen vacancies are unlikely to affect TFT performance metrics such as subthreshold swing or drift mobility. On the other hand, Ga-rich and In-rich oxygen vacancy coordination environments often creates shallow electron trap states that can degrade a-IGZO TFT performance.
\begin{center}
\begin{table}
 \begin{tabular}{c c c c c} 
 Peak& Energy & Density &Dominant V$_O$ cation\\
$\#$ & $\mathrm{eV}$ & $\mathrm{10^{16}cm^{-3}}$ & coordination \\
\hline

0  & 0.12 & 2.0   & \textbf{InInInGa}, InInZn \\
1$^*$  & 0.32  & 9.7   & \textbf{InGaGa} \\
2  & 0.60 & 4.8   & \textbf{InInGaGa}, InInZnZn, InInGaZn, \\
3$^\dagger$  & 0.92 & 0.9   & \textbf{GaGaZn}, GaGaGaIn, InZnZnZn\\
4  & 1.25 & 4.0   & \textbf{InInZnZn}, InInGaZn \\
5  & 1.47 & 26  & \textbf{InInGaZn}, InInGaGa, InInGa, \\
6 & 1.86 & 66  &\textbf{ GaGaInZn}, GaGaZn, GaGaIn\\
7$^\dagger$  & 2.05 & 18  & \textbf{InGaZn}, GaGaInZn, GaGaIn \\
8$^\dagger$  & 2.33 & 270   &\textbf{InGaZnZn}, GaGaInZn, GaGaZnZn\\
9$^*$  & 2.70 & 210   & \textbf{InInGaZn}\\
10$^\dagger$ & 3.03 & 27  & \textbf{InZnZn}, InGaGa, GaZnZnZn \\
\hline
 \end{tabular}

  \caption[example]{\label{table: DFT}
Experimental UP-DoS defect peak energies (CBM = 0) and trap densities for all peaks 0–10. Each peak is matched with dominant oxygen vacancy coordination environments calculated by DFT+U.
\footnotesize $^{\text{*}}$ Specific V$_O$ coordination environment identified, $^\dagger$ Multiple In-poor cation coordination environments possible.}
\end{table}
\end{center}
\par
Figure \ref{fig:DFT}c presents a comparison between experimentally measured DoS and oxygen vacancy defect states as calculated by DFT+U. Black dots correspond to direct numerical differentiation of the raw experimentally obtained trap density spectrum with no smoothing. The colored Gaussian peaks represent the DoS for each measured defect peak as obtained from error function fitting to the same spectrum. Colored lines represent DFT+U calculated oxygen vacancy peaks for a specific cation coordination environment. Many oxygen-vacancy defect peaks with different metal coordination environments overlap in the DoS, allowing multiple possible identifications of the experimental data. The peaks shown in Figure \ref{fig:DFT}c are those that best match experiment and align with the indium enrichment results discussed in the next section. 
Experimental peaks 1 and 9 can be reasonably identified to a single coordination environment due to not having any plausible competing DFT+U peak energy matches and are marked with an asterisk $^*$. For peak 0, 2, 4, 5, and 6, DFT+U cannot assign them to a unique coordination environment, but the dominant metal in each peak can still be identified. Specifically, experimental peaks 0, 2, 4, and 5 are identified as involving coordination environments in which at least 2 neighboring metals are indium, whereas peak 6 is classified as Ga-rich. Finally, peaks 3, 7, 8 and 10 are marked with a dagger $^\dagger$ to indicate multiple DFT+U matches are possible and unambiguous peak identification is not possible. However, Section C  will show these peaks are In-poor. Table \ref{table: DFT} summarizes the DFT+U peak identification results. 
\begin{figure*}
    \includegraphics[width=7 in.]{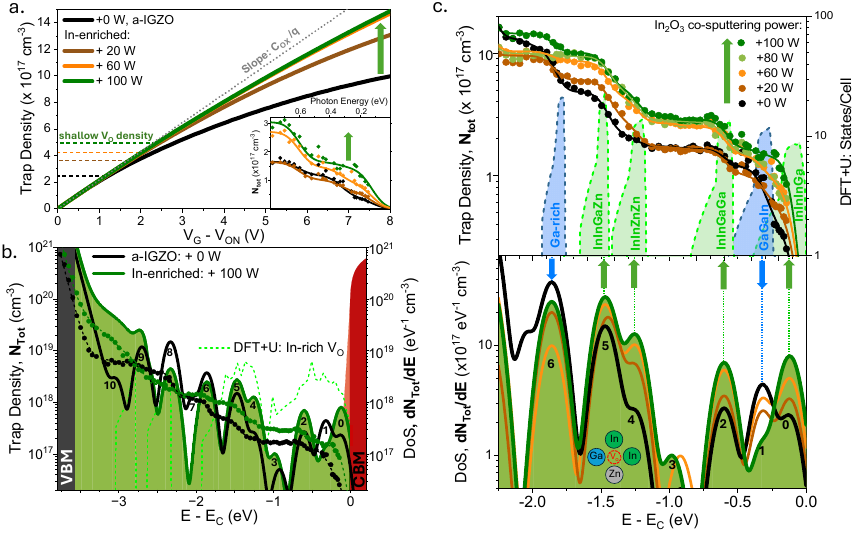}
    \caption{\textbf{(a)} Extracted trap densities versus overvoltage curves for indium-enriched a-IGZO TFTs.
    \textbf{(b)} UP-DoS trap density (dots) and corresponding DoS (filled lines) for an a-IGZO (black) and an In-rich a-IGZO (dark green) TFT. DFT+U simulated DoS for oxygen vacancies with In-rich cation coordination environments (light green). \textbf{(c)} UP-DoS trap density (upper panel) and corresponding DoS (lower panel) of a-IGZO TFTs with different amounts of indium enrichment. Proposed oxygen vacancy cation coordination identifications are indicated (upper panel) and correlated to DoS peaks (lower panel).}
     \label{fig:Indium}
 \end{figure*}
\subsection{a-IGZO Indium Enrichment DoS Trends}
The objective of this section is to experimentally identify which oxygen vacancy peaks involve indium-rich cation coordination environments. Indium enrichment is accomplished via co-sputtering from an indium oxide target.
\par
Figure \ref{fig:Indium}a presents transfer curve extracted trap density (volumetric, Q$_{t}($V$_G$)$/q\times$d, d$\approx$10 nm) versus overvoltage plots for a-IGZO TFTs with different amounts of indium incorporated into the channel. The extracted trap density increases with increasing co-sputtering power and appears to saturate for a co-sputtering power above $\sim$ 60 W. Thus, in this experiment, a-IGZO quality, in terms of shallow trap density, degrades with increased indium incorporation. This is expected since process optimization of co-sputtered TFTs has not been undertaken. Importantly, the purpose of In-enrichment is to aid in oxygen vacancy peak assignment without introducing interfering effects from process optimization.
\par
In Figure \ref{fig:Indium}b, the UP-DoS trap density and corresponding DoS are compared across the full bandgap for the 0 W (black, no co-sputtering) and 100 W (green, maximum Indium incorporation) a-IGZO TFTs. In the most In-rich TFT, an additional feature near the valence band appears at $\sim$3.0 eV with an energy and lineshape closely matching the indirect bandgap of a possible In$_2$O$_3$ phase. \cite{king2009band,bierwagen2015indium} Comparing the total experimental trap density just below this bandgap feature, the total oxygen vacancy trap density (shallow and deep) remains approximately equal between the two samples, while the amplitudes of individual defect peaks vary by $\sim$2-8$\times$ after In-enrichment. In the upper portion of the bandgap (for experimental peak number 5, 4, 3, 2 and 0), the trap density of In-rich a-IGZO increases significantly. This observation is in strong agreement with the overlaid DFT+U DoS (light green), which shows a large increase in the DoS of In-rich oxygen vacancy states in this energy region. In the midgap, experimental peaks numbered 6, 7, and 8 are all observed to decrease with In-growth power, consistent with a Ga- or Zn-rich oxygen vacancy coordination environments. This assignment is further supported by DFT+U simulations, which predict few or no In-rich defect peaks in this region relative to the total. Finally, near the valence band edge, experimental peak 9 increases with indium incorporation, consistent with DFT+U calculating an InInGaZn coordinated oxygen vacancy at the corresponding energy.
\par
From the perspective of TFT electrical performance, the most shallow oxygen vacancy traps are of primary interest. Figure \ref{fig:Indium}c shows the UP-DoS trap density (upper panel) and DoS (lower panel) for this upper portion of the bandgap for co-sputtering powers 0-100 W. In the upper panel, DFT+U simulated oxygen vacancy peaks with specific cation coordination environments are proposed. Defect peak numbers 0, 2, 4, 5 increase systematically with In-incorporation, and are attributed to In-rich oxygen vacancy coordination environments. Conversely, defect peak numbers 1 and 6 decrease with In-incorporation, consistent with a Zn- or Ga-rich oxygen vacancy coordination.  This further confirms shallow peak 1 as a likely GaGaIn 3-atom coordinated oxygen vacancy. This GaGaIn coordinated oxygen vacancy is observed to be the dominant trap that defines subthreshold behavior and degradation characteristics of well-functioning a-IGZO TFTs (see Fig \ref{fig:simulation}). To a large extent, process optimization of a In:Ga:Zn = 1:1:1 a-IGZO TFT involves reducing the density of this Ga-rich peak 1 as much as possible. Defect peak number 0 is attributed to an In-rich coordinated oxygen vacancy environment (InInInGa or InInZn). This peak is generally observed to have low density in UP-DoS assessment of 1:1:1 a-IGZO TFTs. For optimization towards a well-functioning indium-rich a-IGZO TFT, this defect peak must be suppressed.
\section{Conclusions}
The subgap density of states of a-IGZO TFTs are measured using tunable lasers from the valence band to within 0.14 eV of the conduction band mobility edge by ultrabroadband photoconduction density of states (UP-DoS) microscopy. First-principles device physics simulation using the experimentally obtained defect DoS, together with statistical analysis of 25 TFTs fabricated under varying processing conditions, demonstrates that shallow oxygen vacancy defect density critically controls TFT metrics such as subthreshold swing, threshold voltage, and drift mobility. For a set of TFTs with varying processing conditions, the simulation from UP-DoS data accurately reproduces the subthreshold TFT transfer curve with no adjustable parameters. Simulating the entire TFT transfer curve requires only a single adjustable parameter: the conduction band tail Urbach energy. Alternatively, the shallow trap DoS responsible for subthreshold TFT behavior can instead be extracted from a TFT transfer curve, and independently agree with trap density trends measured directly by UP-DoS.
\par
An indium enrichment study of a-IGZO TFTs fabricated via co-sputtering, in conjunction with DFT+U simulations, facilitates the proposal of specific cation coordination for the 11 oxygen vacancy peaks observed with UP-DoS. In conventional 1:1:1 ratio a-IGZO TFTs, the dominant shallow trap state determining TFT performances is centered at $\sim$0.32 eV from the conduction band mobility edge and is ascribed to a Ga-Ga-In metal-coordinated oxygen vacancy. For an indium-rich a-IGZO TFT, a more shallow trap state appears at $\sim$ 0.12 eV below the conduction band mobility edge and is attributed to an InInInGa or InInZn metal-coordinated oxygen vacancy. Although indium enrichment of an amorphous oxide semiconductor is an attractive approach for achieving higher TFT mobility, due to a corresponding decrease in the electron effective mass, process optimization of such devices is likely challenging. This challenge arises from the need to suppress shallow In-rich oxygen-vacancy defect states that accompany such indium enrichment.
\section{Experimental Section}
\par
\medskip
\subsection{Preparation of a-IGZO TFTs}
A diverse selection of a-IGZO TFTs is prepared for robust demonstration of how well UP-DoS simulates transfer curves that include: (1) back-gated TFTs from designs appropriate for display panel applications, (2) nanoscale dimension top-gated TFTs for DRAM memory application developments, and (3) back-gated TFTs with systematic In-enrichment.  More specific details are available in the Supplementary Information Section S1.
\subsection{Ultrabroadband Photoconductance Density of States (UP-DoS) Microscopy of a-IGZO TFTs}
\label{sec:UP-DoS}
The integrated trap density, N$_\text{tot}$, of a-IGZO TFTs is measured using Ultrabroadband Photoconduction Density of States (UP-DoS) microscopy that has been modified to selectively excite shallow subgap defects within 0.14 eV of the conduction band mobility edge.\cite{vogt2025ultrabroadband} The inset of Fig. \ref{fig:2} displays a rough schematic of the UP-DoS setup where the incident laser directly illuminates the TFT active channel using a reflective objective (52x, NA=0.65) to create a near-diffraction-limited laser spot.  Multiple tunable femtosecond lasers are coupled into the UP-DoS microscope via a home-built 4f-confocal piezo-scanning mirror setup that enables continuous, selective laser excitation of subgap defect states. Collectively, the UP-DoS setup achieves full spectral (0.14 to 5 eV) and spatial resolution of the photon-normalized TFT photoconductance response (See Figure S4 in Supporting Information).\cite{vogt2020ultrabroadband,vogt2025ultrabroadband}

In UP-DoS, the UV-visible range is covered using a Ti:Sapphire laser with harmonics extender (Coherent Chameleon Ultra II and APE Harmonix), and the near-IR, 0.3-1.2 eV range is achieved using an automated optical parametric oscillator (OPO, APE Compact).   Finally, this work enables a direct probe of shallow defect states near the conduction band mobility edge.  This innovation is achieved through home-built difference-frequency generation (DFG) mixing of the OPO signal and idler laser lines on a HgGa$_2$S$_4$ nonlinear crystal, which provides tunable IR-laser energies down to $\sim$0.1 eV.\cite{beutler2016femtosecond} 

The UP-DoS microscopy method employs a lock-in amplifier (Zurich MFLI) with a 585 Hz optical chopper frequency (Thorlabs MC2000B) to enhance the signal-to-noise ratio and largely remove transfer-curve drift effects that may otherwise affect the measured TFT photoconductance.   Collection is via RF-probes in series with a current preamplifier (Femto DDPCA-300) at a fixed positive bias.   The photoconduction signal is per photon normalized, such that $I_{\text{norm}}(h \nu)= h\nu I_{PC}/P$, where $I_{PC}$ is the recorded increase in conduction due to the incident light and P is the incident power (nW to $\mu$W regime). UP-DoS is conducted at a fixed gate voltage with the TFT turned on and operating in the linear regime.  Under these approximations,  the TFT photoconduction is proportional to the number of carriers excited from the defect state to the conduction band by the equation: \cite{vogt2020ultrabroadband,mattsson2025defect}
\begin{equation}
    N_{tot}(h \nu)=I_{\text{norm}}(h\nu)\left[ \frac{C_{ox}k_{0}}{qd}\left( \frac{\partial I_{D}}{\partial V_{G}}\right)^{-1} \right]
    \label{eqn:ubpc}
\end{equation}
where $k_{0}$ = $\frac{P_{sat}}{h\nu_{E_g}}$ is the power saturation constant, $\frac{\partial I_{D}}{\partial V_G}$ is the transfer curve slope and $d$ is the accumulation layer thickness.   For subgap photoconduction in n-type TFTs, the photoconduction response $I_{\text{norm}}(h\nu)$ corresponds to the optical excitation from all filled trap states within energy $h\nu$ of the conduction band minimum. As such, the raw UP-DoS spectrum, $I_{\text{norm}}(h\nu)$ is directly proportional to the trap density, $N_{tot}(h \nu)$, and in turn the subgap DoS as $ \mathrm{DoS}(E_C-E) = \dfrac{dN_{tot}(h\nu)}{dh\nu}$. Since smoothing involved in numerical derivatives can be subjective, as shown in Fig. 2a, the subgap DoS is instead retrieved by fitting $N_{tot}(h \nu)$ to a sequence of error functions offset by each defect energy to match the expected stepwise decreasing photoconduction response. \cite{mattsson2025defect} This analysis approach is compared with numerical derivatives of the UP-DoS spectra obtained with minimal or no smoothing. This comparison, along with further experimental details, are shown in Supplementary Information Section S4.
\par
\subsection{Simulation of TFT Transfer Curves from Experimental DoS}
\label{sec:sim}
The drift mobility of a-IGZO TFTs is simulated from experimentally measured DoS by Fermi-Dirac statistics. The trapped and free charge sheet densities, induced by an applied gate voltage, Q(F$_n$) and Q$_T$(F$_n$), are evaluated as a function of the electron quasi-Fermi energy level, F$_n$.\cite{wager2022amorphous} Once Q(F$_n$) and Q$_T$($F_n$) are known, the simulated drift mobility, $\mu_{SIM}$(F$_n$) is calculated as:\cite{nenashev2019percolation,wager2022amorphous}
\begin{equation}
    \mu_{SIM}(F_n)=\mu_0  \left( \frac{Q(F_n)}{Q(F_n)+Q_T(F_n)}\right)
    \label{eqn:trap}
\end{equation}
where $\mu_0$ is the trap-free extended state mobility. All simulations assume a $\mu_0$ = 23 cm$^2$ $V^{-1}$ $s^{-1}$, which is extracted from a best fit to Equation \ref{eqn:trap} of the empirical mobility data of Figure \ref{fig:2}b and is consistent with previous literature estimates \cite{fishchuk2016interplay,stewart2016amorphous,stewart2016thin,wager2022amorphous} Note the trapped and free carrier densities are simply n$_T$ = Q$_T$/q and n = Q/q, where each electron has a charge of q. To relate the quasi-Fermi energy level, F$_n$ to the specific applied gate voltage, V$_G$, the charge sheet approximation, $C_{ox}(V_G-V_{ON})=Q(F_n)+Q_T(F_n)$ is used to plot the DoS, and related simulated quantities on a V$_G$ axis (e.g. see Figs. 1b and 4b-d).\cite{wager2022amorphous} 

\par To convert simulated drift mobility, $\mu_{SIM}$(V$_G$), to a corresponding simulated TFT source-drain current, the following EKV drain-current model is used: \cite{enz1995analytical}
\begin{multline}
I_{D,SIM} = \mu_{SIM} \times2 \left( \frac{k_B T}{q} \right)^2 C_{ox} \frac{W}{L}
\\
\Bigg\{
\left[
\ln \left( 1 + \exp \left( \frac{q (V_G - V_{ON})}{2 k_B T} \right) \right)
\right]^2 \\
-
\left[
\ln \left( 1 + \exp \left( \frac{q (V_G - V_{ON} - V_D)}{2 k_B T} \right) \right)
\right]^2
\Bigg\}
\end{multline}
The above equation is further used to extract an experimental TFT drift mobility, $\mu_{EKV}$, directly from a transfer curve by setting the I$_D$ factor equal to the experimentally measured source-drain current and subsequently solving for the mobility factor. \cite{wager2022amorphous,wager2022thin} To convert from $\mu_{EKV}(V_G)$ to the experimental free and trapped charge density induced by the applied gate voltage, we again employ the charge-sheet approximation to obtain,\cite{wager2022thin}
\begin{align}
Q_T(V_G)  &= C_{ox} (V_G - V_{ON}) \left[ 1 - \frac{\mu_{\text{EKV}}(V_G)}{\mu_0} \right]
\\
Q(V_G)  &= C_{ox} (V_G - V_{ON})\left[  \frac{\mu_{\text{EKV}}(V_G)}{\mu_0} \right]
\label{eqn:freedensity}
\end{align}
where $\mu_{EKV}$ is experimentally extracted drift mobility from the transfer curve and $\mu_0$ is the same free extended state mobility of 23 cm$^2$ V$^{-1}$ s$^{-1}$. In the density of states simulation, the off-state quasi-Fermi energy level, at which V$_G$ = V$_{ON}$, is set such that the simulated Q($F_n$) = Q($V_G=V_{ON}$) equals the extracted free charge density at V$_G$ = V$_{ON}$ by Eq. \ref{eqn:freedensity}. This off-state quasi-Fermi level marks the boundary beyond which deeper states have negligible influence on gate voltage-induced electron trapping and is positioned $\approx$ 0.65 eV below the conduction band mobility edge. For TFT simulation purposes, the absolute scaling of the measured density of states is initially treated as an adjustable parameter, while the relative scaling between different TFTs remains fixed and is held constant. For more details of the TFT simulations, together with a full list of equations and Fermi-Dirac statistics used, please refer to Supplementary Information Section S7 and the Supporting Code provided.

\subsection{DFT+U Defect Simulations}
The DFT+U methods employed a multi-stage approach that starts by placing twenty formula units of IGZO randomly into a cubic lattice cell with 1.3x more volume than the IGZO crystal structure. This was proceeded by taking the randomized cubic cell through a melt and quench process by employing the General Lattice Utility Program (GULP), similiar to previous simulation studies.\cite{vogt2020ultrabroadband,Gale1997,Buchanan2017} The quenched amorphous structures were further refined using Vienna ab initio Simulation Package (VASP) using the Perdew-Burke-Ernzerhof (PBE) generalized gradient approximation (GGA) functional theory, which resulted in amorphous IGZO structures with an approximate density of $\sim$6.1 g/cm$^3$ that is consistent with results reported by Kamiya et al. \cite{Kamiya2010, kresse1996efficiency, kresse1996efficient, kresse1999ultrasoft,perdew1996generalized, kresse1993ab} To introduce an oxygen vacancy to the amorphous structure, a single oxygen atom is randomly removed before the cell is relaxed. The local coordination environment of each simulated oxygen vacancy is defined as the set of atoms within 2.5 Å of the removed oxygen atom. More complete details on the DFT+U simulation, including the cell-relaxation approach and scissoring method, are provided in Supplementary Information Section S3. 
\medskip
\par
\textbf{Supplementary Information} \par Details on sample characteristics, device simulations, DFT+U simulations and optical conductivity data modeling methods.
\medskip

\textbf{Acknowledgments} \par This work is supported by the SAMSUNG Global Research Outreach (GRO) Award and partly by Samsung Electronics Co., Ltd. (IO250414-12603-01) and a NSF Grant (DMR-1920368).


\textbf{Competing interests:} M.M., K.V. and M.G. are named inventors on a U.S. patent application (US 2025/0102448 A1) covering aspects of the ultrabroadband photoconduction methodology described in this work, which is assigned to Oregon State University.
\medskip

\textbf{Data Availability Statement}: The data that support the findings of this study are available from the corresponding author upon request. 

\medskip


%

\bibliographystyle{MSP}

\bibliography{mybib.bib}   

\pagebreak

  
  
\end{document}